\documentclass[amssymb,prb,twocolumn,floats,amsmath,showpacs]{revtex4}
\usepackage[T1]{fontenc}
\usepackage{bm}
\usepackage{graphicx}
\usepackage{amssymb}
\usepackage{leftidx}
\usepackage{upgreek}
\usepackage{siunitx}
\usepackage{amsfonts}
\usepackage{amsmath} 
\usepackage[outdir=./]{epstopdf}
\usepackage{color}

\begin{document}

\title{Origin of luminescence quenching in structures containing CdSe/ZnSe quantum dots with a few Mn$^{2+}$ ions}

\author{K.~Oreszczuk}
\email{Kacper.Oreszczuk@fuw.edu.pl}
\author{M.~Goryca}
\author{W.~Pacuski}
\author{T.~Smole\'nski}
\author{M.~Nawrocki}
\author{P.~Kossacki}

\affiliation{Institute of Experimental Physics, Faculty of Physics, University
of Warsaw, ul. Pasteura 5, PL-02-093 Warszawa, Poland}


\begin{abstract}
We present a detailed spectroscopic study of the photoluminescence quenching in an epitaxial structures containing CdSe/ZnSe quantum dots doped with low concentration of Mn$^{2+}$ ions. Our time-resolved and time-integrated experiments reveal the origin of the quenching observed in macro-photoluminescence studies of ensembles of such dots. We show that incorporation of even a few ions to an individual dot does not quench its luminescence effectively, although some fingerprints of expected spin-dependent quenching are visible. At the same time, the presence of Mn$^{2+}$ ions in the sample significantly affects the luminescence intensity of the wetting layer, resulting in a quenching of the global luminescence from studied structure. On the other hand, the luminescence decay dynamics is found to be independent of the presence of Mn$^{2+}$ ions, which suggests that the observed quenching occurs for the excited excitonic states.
\end{abstract}


\maketitle

\section{Introduction}

Owing to the exchange interaction between magnetic ions and band carriers, Diluted Magnetic Semiconductors (DMS) exhibit giant Zeeman effect and related extraordinary magneto-optical properties\cite{kossutdmsbook}. Influence of the magnetic ions can be further enhanced by introducing them to low dimensional semiconductor structures,\cite{Gaj1994,wu1997, kratzert2001, oka2002, Boukari2002,cdtemn,Beaulac_S09} which could be applied in the field of spintronics, \cite{Dietl2014} solotronics,\cite{Flatte2011,nature2014} and in the photophysics\cite{peng2012}. From this point of view, particularly interesting are structures constructed of Mn-doped wide-gap II-VI semiconductors, such as ZnSe, ZnS or ZnO. This is because excitons in wide-gap semiconductors exhibit high binding energy and high oscillator strength, which open a possibility of room-temperature operation, and result in high efficiency of photon sources as well as strong coupling with photonic modes \cite{Fedorych2012,Sebald2012,Lai2013}. However, the suppression of excitonic luminescence resulting from addition of Mn$^{2+}$ ions to both bulk material and to the low dimensional structures was observed \cite{bhargava1994, falk2002, Pacuski2011, galkowski2015}. This is due to the opening of new, sensitive to the magnetic field, channels of non-radiative cross-relaxation of excitons leading to the excitation of the Mn$^{2+}$ ions \cite{nawrocki, kim2000, agekyan2002, Lee2005, beaulac2008, chernenko2010, peng2012, chernenko2015}. Therefore, of great interest were the results obtained on recently grown single self-organized CdSe/ZnSe quantum dots (QDs) containing individual Mn$^{2+}$ ions, which indicated that in such structures a suppression of the luminescence does not occur effectively, while the specific magneto-optical properties of DMS are preserved \cite{nature2014,smolenski2015,appl2015}.

Explanation of the absence of an effective luminescence quenching in a QD with a single magnetic ion may have a double importance. On the one hand to understand the physics which lies behind it, on the other it may indicate how to reduce luminescence quenching in structures with a higher content of Mn$^{2+}$, which from the point of view of their magneto-optical properties would be highly desirable. For this purpose in the present study we investigate the luminescence quenching in the single CdSe/ZnSe QDs with both individual and several  Mn$^{2+}$ ions. We demonstrate that in both cases the luminescence quenching is not effective, but a fingerprint of quenching can be deduced from comparison of optical lines related to various spin states of an individual Mn$^{2+}$ in a QD. We also show that the strong quenching observed in the measurements performed without spatial resolution has its origin in the wetting layer.

\section{Sample and experimental setup}

\begin{figure}
\includegraphics[width=85mm]{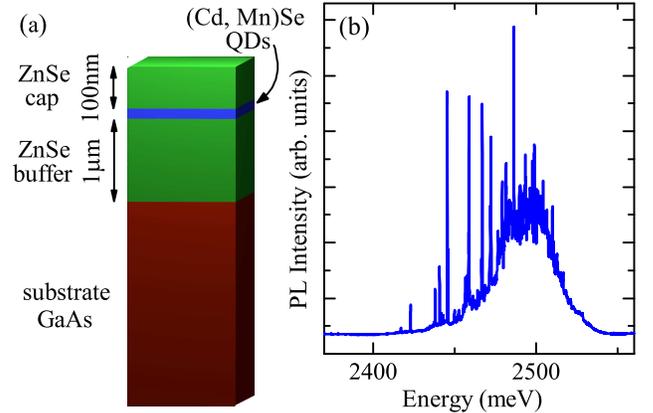}
\caption{ (a) Cross-section scheme of the studied samples containing CdSe/ZnSe QDs with Mn$^{2+}$ ions (not in scale). (b) Micro-photoluminescence spectrum of CdSe/ZnSe QDs measured at randomly selected position on the sample surface.
}
\label{probka}
\end{figure}

The samples studied in this work contain a single layer of self-assembled  CdSe/ZnSe QDs, some of them doped with manganese ions. They are grown using molecular beam epitaxy (MBE) on GaAs (100) substrate. The growth procedure is particularly simple. First, we grow \SI{1}{\upmu m} ZnSe buffer layer, then 2 monolayers of Cd$_{1-x}$Mn$_x$Se, which are immediately transformed into QDs, and finally \SI{100}{nm} of ZnSe cap (see Fig \ref{probka}(a)). The concentration of dopants was different for various studied samples. In the case of the sample being optimized for the measurements of QDs with single Mn$^{2+}$ ions, the dopant concentration was similar to concentrations in other studies of such dots \cite{cdtemn,inasmn,miganieco,nature2014,Varghese2014,nature2016,Lafuente2016,piwowar2016,Fainblat2016,Smolenski2017prb} as well as in the studies of dots with two Mn$^{2+}$ ions\cite{Besombes2012, Krebs2013}. On the other hand, the sample optimized for the measurements of QDs with a few Mn$^{2+}$ was doped with a few times higher amount of dopants, but still smaller as compared to the studies of QDs with many of Mn$^{2+}$ ions\cite{Wojnar2007,Clement2010,Klopotowski2011}. The nominal magnetic-dopant concentrations in Cd$_{1-x}$Mn$_x$Se formation layer of both aforementioned samples are $x=1\%$ and $x=3\%$, respectively. A sample without Mn was also grown for the reference purposes.

The identification of the QDs with various number of magnetic ions as well as all further measurements were performed at temperature of about \SI{1.6}{K} in liquid-helium-cooled cryostat. The cryostat was equipped with a superconducting magnet, which produced magnetic fields of up to \SI{10}{T}. If not stated otherwise, the time-integrated photoluminescence (PL) measurements were performed with the use of continuous-wave (CW) laser with wavelength equal to \SI{488}{nm} (\SI{2540}{meV}), which is below the barrier bandgap (\SI{2820}{meV}). For micro-photoluminescence ($\mu$-PL) measurements the sample was attached directly to immersion reflection microscope. In this configuration the size of the laser spot on the sample surface was below 1~$\upmu$m. Due to the high density of typical self-organized QDs, even for relatively small laser spot size a large number of dots are optically excited, although it is possible to resolve the luminescence of an individual QD in the low-energy tail of the PL band (see Fig \ref{probka}(b), sample with lower Mn content).

The luminescence signal was spectrally resolved by a \SI{750}{mm} focal length monochromator equipped with a CCD camera. For time-resolved measurements, a streak camera with a \SI{425}{nm} (\SI{2916}{meV}) pulsed laser were used. Time-integrated PL spectrum and its evolution in the magnetic field exhibited no qualitative dependence on the type of the exciting laser, neither in the micro-PL nor in the macro-PL measurements. This includes both the excitation with the lasers mentioned earlier and with an above-the-barrier \SI{405}{nm} (\SI{3060}{meV}) CW laser.

\section{Photoluminescence quenching - ensemble measurements}

\begin{figure}
\includegraphics[width=85mm]{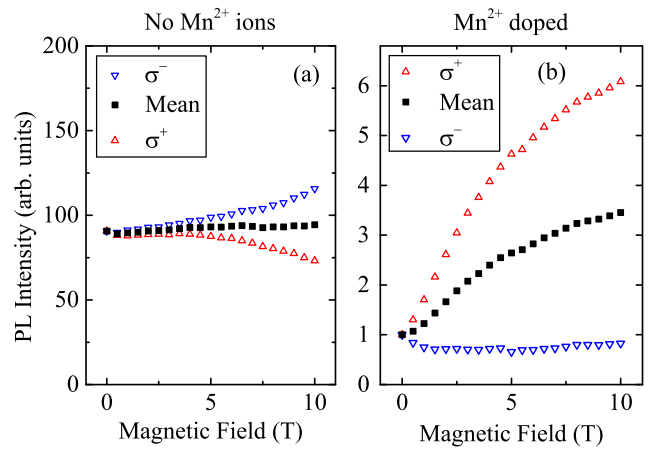}
\caption{Dependence of the integrated PL intensity of an ensemble of QDs measured in $\sigma^{+}$ polarization of detection as a function of the magnetic field for (a) the sample without Mn ions and (b) the sample with QDs with a few Mn$^{2+}$ ions. In the latter case significant rise of the PL intensity with magnetic field is observed, which is typical for magnetic-ion-related PL quenching. Excitation power is equal to \SI{35}{\upmu W}. In both plots the PL intensity is normalized to the PL intensity of doped sample at \SI{0}{T}.}
\label{mocbuly}
\end{figure}

In order to verify the presence of the PL quenching in our highly Mn-doped sample, we performed macro-PL measurements of QDs ensemble from this sample and from the reference sample without Mn dopants. As shown in Fig. \ref{mocbuly}, in the case of Mn-doped sample a significant rise of the PL intensity in $\upsigma^{+}$ circular polarization of detection is observed. Such behavior is a well-known fingerprint of magnetic-ions-related PL quenching \cite{nawrocki,oka,Lee2005,lee2007,beaulac2008,Crooker2011PRL} due to non-radiative exciton recombination channel corresponding to excitation of the Mn$^{2+}$ ions. The total spin $S$ of the ion is equal to~$\frac{5}{2}$ for $\leftidx{^6}{A_1}$ ground state, while for the excited $\leftidx{^4}{T_1}$ and $\leftidx{^4}{T_2}$ states the spin~$S=3/2$. The selection rules for the spin projection forbid the excitation of an ion occupying its ground states with $S_z = \pm \frac{5}{2}$. Thus, the spin orientation towards these states in high magnetic fields leads to a suppression of the PL quenching. The effect is less pronounced for high excitation power, probably due to increase of the effective temperature of the Mn$^{2+}$ ions. Importantly, such rise of the PL intensity is not observed in similar experiment carried out on the sample without Mn$^{2+}$ ions. In this case, the only observed variation of the PL intensity is a slight polarization of the whole spectrum due to the excitonic spin relaxation towards lower-energy states\cite{johnston2001,tsitsishvili2003}. 

\begin{figure}
\includegraphics[width=85mm]{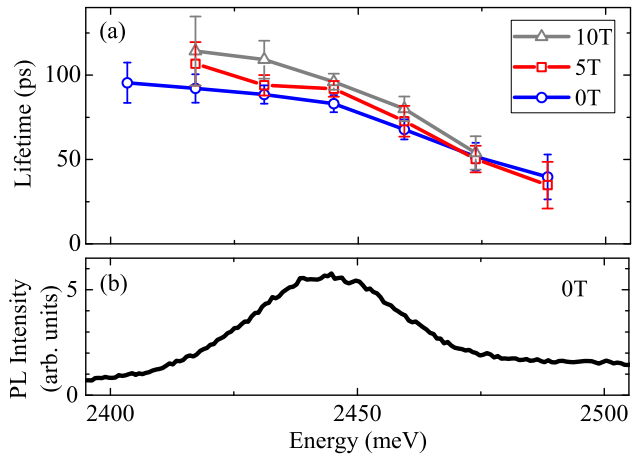}
\caption{(a) Lifetime of the ensemble luminescence from the sample with QDs with a few Mn$^{2+}$ ions in various magnetic fields, plotted as a function of the photon energy. The presented values were obtained by fitting the measured PL decay profiles with mono-exponential curves. (b) Time-integrated luminescence spectrum of the QDs excited by a pulsed laser.}
\label{czaszyciabuly}
\end{figure}

To confirm the nature of the observed phenomena in the Mn-doped sample, the measurements of the luminescence lifetime were performed. One could expect the lifetime to rise significantly with magnetic field, as the non-radiative recombination channel related to the PL quenching by Mn$^{2+}$ ions is progressively turned off, but this is not the case in our sample. In fact, the average lifetime is found to be equal to about \SI{100}{ps} independently of the value of the magnetic field (see Fig. \ref{czaszyciabuly}). This finding indicates that the observed PL quenching is not related to the non-radiative recombination channels of the relaxed exciton states (i.e., the states to which the PL is related), but must occur before the relaxation of photo-created carriers. The decrease of the decay time with the increase of the PL energy is most probably related to the thermal relaxation of the carriers\cite{Korona2005}.

\section{Identification of single QD\lowercase{s} with a few
M\lowercase{n}$^{2+}$ ions}

In order to investigate in detail the origin of the observed PL quenching, we performed micro-PL measurements of single QDs. Due to small laser spot on the sample it was possible to separate well-resolved PL lines originating from single QDs in the low-energy tail of the QD ensemble spectrum. Those lines appear on the top of spectrally-broad background typically ascribed to the wetting layer, which is usually present in the samples with QDs grown in Stranski--Krastanov mode \cite{makino2003}. Among typical sharp lines related to non-magnetic dots, we also observe broader features of similar, triangular shape with well visible splitting in the middle (see Fig. \ref{symulacje0T}(g-i)). The typical width of such features being identified in our samples is of about \SI{1.2}{meV}, while the width of the middle-energy splitting is typically equal to about \SI{0.4}{meV}. Since such features are not observed in the case of similar samples with no magnetic impurities, we tentatively relate them to QDs containing a few Mn$^{2+}$ ions. Inferring from the width of the features, the Mn$^{2+}$ ions are positioned in the QD in such a way that the exciton-Mn exchange interaction has a modest strength when compared to the case of previously investigated CdSe QDs with single Mn$^{2+}$ ions \cite{nature2014, appl2015}. In order to support this hypothesis we perform simulations of the exciton spectrum in QDs with different number of Mn$^{2+}$ ions. The model used in such simulations is based on the model of the exciton in QDs with single dopants\cite{nature2014}, and further expanded to describe interactions with many ions. The Hamiltonian of the exciton in such a QD is given by:
\begin{figure}
\includegraphics[width=85mm]{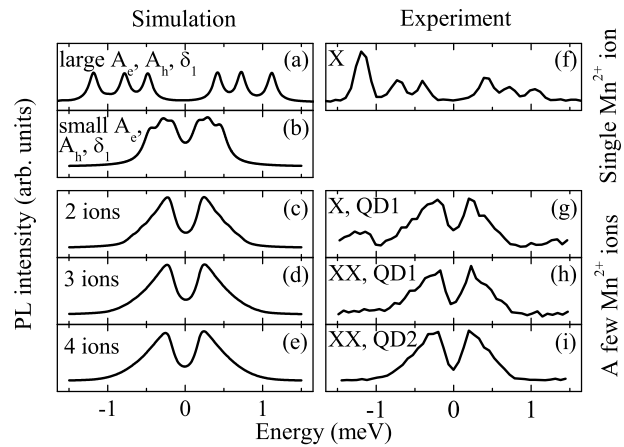}
\caption{Left -- simulations: (a) Spectrum of QD with single Mn$^{2+}$ ion. (b) QD with single Mn$^{2+}$ ion, low exchange constant for exciton and ion. (c-e) QDs with two, three and four Mn$^{2+}$ ions, respectively. Right -- experiment: (f)  QD with single Mn$^{2+}$ ion. (g), (h) exciton and biexciton in QD with a few Mn$^{2+}$ ions. (i) biexciton in QD with a few Mn$^{2+}$ ions.}
\label{symulacje0T}
\end{figure}
\begin{figure}
\includegraphics[width=85mm]{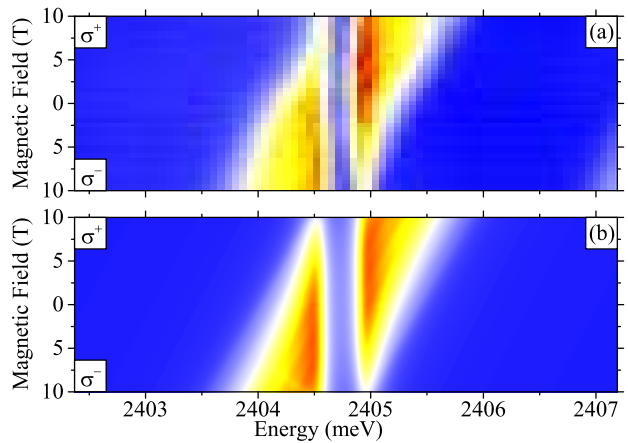}
\caption{PL spectrum of QD with a few Mn$^{2+}$ ions plotted as a function of the magnetic field (a) and a corresponding simulation (b).}
\label{mapaisymulacja}
\end{figure}
\begin{equation}\label{H}
\begin{split}
  H_{X}= & E_0+\Delta_{l-h} \frac{1}{2}[(3/2)^2-(J_h^z)^2]-\delta_0
\frac{2}{3} J_e^z J_h^z+\\
        + &\delta_1 \frac{2}{3}\left[J_e^y \left(J_h^x\right){}^3-J_e^x
\left(J_h^y\right){}^3\right]+\\
        - & \mu_B B (g_e J_e^z+g_h J_h^z)+\\
        +& \sum_k (-\mu_B B g_m S_k^z+ A_{h_k} \vec{J}_{h} \vec{S}_{k}+ A_{e_k}
\vec{J}_{e} \vec{S}_{k}),
        \end{split}
\end{equation}
where $J_e$, $J_h$, $S_k$, are the spin operators of electron, hole, and \textit{k}th Mn$^{2+}$ ion spin operators. Light-heavy hole energy difference is denoted by $\Delta_{l-h}$, $\delta_0$ represents the dark-bright exciton splitting, while $\delta_1$ corresponds to the splitting of the two bright excitons due to the anisotropic part of the electron-hole exchange interaction. $g_e$, $g_h$, $g_m$ are the g-factors of the electron, hole, and Mn$^{2+}$ ions, respectively. $E_0$, $\delta_1$, $g_e$, $A_{e_k}$ were chosen to fit the experimental data. Ratios $g_e/g_h=1.2$ and $A_{e_k}/A_{h_k}=-1/2$ as well as the values of $\delta_0=\SI{2.5}{meV}$, $g_m=2$ and $\Delta_{l-h}=\SI{30}{meV}$ are based on the results of previous studies of Mn-doped CdSe QDs\cite{nature2014, smolenski2015}.

Results of those simulations are illustrated in Fig. \ref{symulacje0T}, together with experimentally obtained spectra of QDs containing one and -- presumably -- a few magnetic impurities. Even though we are not able to resolve the multitude of excitonic lines in the case of a few Mn$^{2+}$ ions, it is still possible to analyze general shape of the resulting broad feature in the spectrum. For example, the splitting always visible in the middle of the feature is related to the anisotropic part of the electron-hole exchange interaction ($\delta_1$). Importantly, we can distinguish QDs with exactly one Mn$^{2+}$ ion from those containing more dopants. In the case of a single dopant, the broad feature in the spectrum maintains approximately rectangular shape for a broad range of QD parameters, i.e., anisotropy splitting, g-factors of the exciton and the ion or the exchange constants (see Fig. \ref{symulacje0T}(a-b)). The identification of the QD with a few Mn$^{2+}$ dopants may be further confirmed by the analysis of its evolution in a magnetic field applied in the Faraday geometry. As shown in Fig. \ref{mapaisymulacja}, such evolution is well reproduced by our theoretical model.

\section{Photoluminescence quenching in an individual QD}

\begin{figure}
\includegraphics[width=85mm]{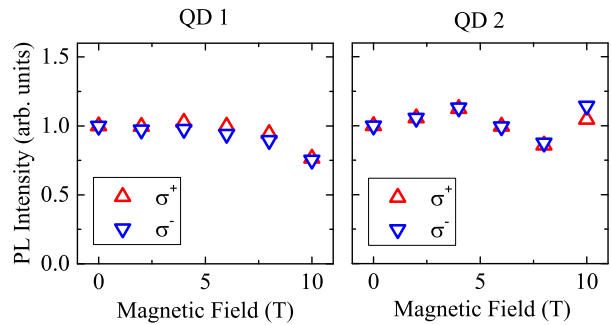}
\caption{Dependence of integrated PL intensity of two QDs with a few Mn$^{2+}$ ions. Intensity is measured as a function of magnetic field in both polarizations of detection.}
\label{mockropki}
\end{figure}

\begin{figure}
\includegraphics[width=85mm]{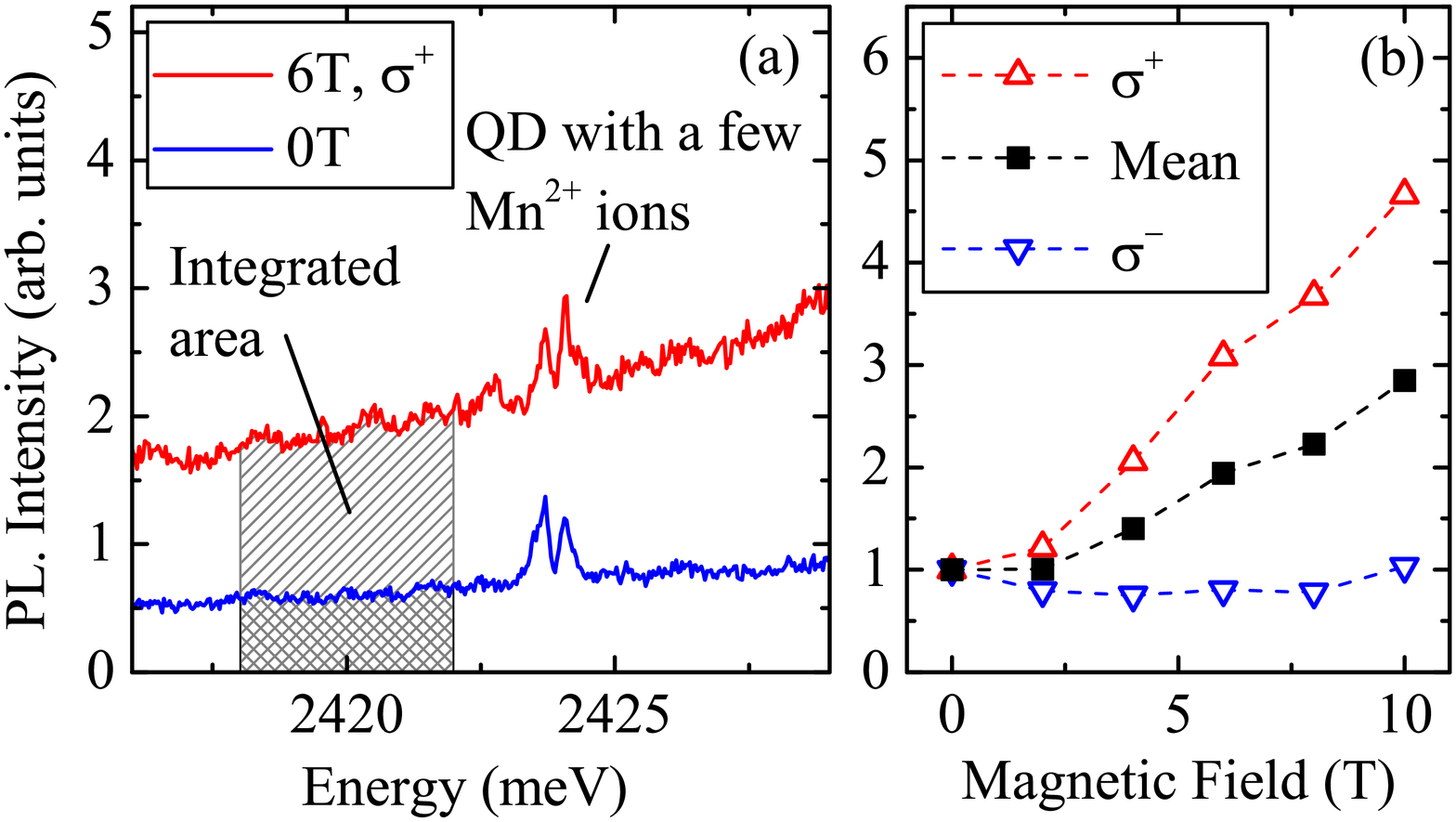}
\caption{(a) PL spectrum of the QD with a few Mn$^{2+}$ ions and the broad background as a function of magnetic field. (b) Integrated PL Intensity of the broad background photoluminescence as a function of magnetic field in both circular polarizations of detection. Excitation power is equal to \SI{100}{\upmu W}.}
\label{mikrotlo}
\end{figure}

The existence of the quenching of the excitons confined inside QD with a few Mn$^{2+}$ ions was verified with the use of two different methods. The first one is based on measuring the total intensity of the PL lines originating from the recombination of excitons confined in the QD for various values of the magnetic field applied in the Faraday configuration. The sample was illuminated with a \SI{488}{nm} CW laser and the intensity of the PL lines was determined after subtracting the broad background from the PL spectra. If the effective PL quenching was present in the observed QDs, the PL intensity should significantly increase with the magnetic field due to the blockade of non-radiative recombination channels \cite{nawrocki, Lee2005}. In our case we observed no significant variation of the photoluminescence intensity for all magnetic fields up to \SI{10}{T} and both circular polarizations of detection (see Fig. \ref{mockropki}). Such a variation was, however, observed for the broad background (see Fig. \ref{mikrotlo}) related to the wetting layer.

The second method is based on a direct measurement of the PL lifetime. Recombination channel corresponding to intraionic transitions -- if such exists -- should shorten the lifetime of the excitons confined in the QD. The presence of the PL quenching can be verified either by analyzing dependence of luminescence lifetime as a function of the magnetic field or by a direct comparison of the lifetime of the exciton in a QD without Mn$^{2+}$ ions and in a QD with a few Mn$^{2+}$ ions. We have measured lifetimes of exciton (X) and biexciton (XX) luminescence peaks as a function of the magnetic field. Neither of these two lifetimes exhibited significant dependence on the field. For an example QD with a few Mn$^{2+}$ ions, the X lifetime is equal to \SI{250}{ps} $\pm$ \SI{15}{ps}, while the XX lifetime is \SI{120}{ps} $\pm$ \SI{7}{ps} for all magnetic fields up to \SI{10}{T}. Also, the lifetimes of both excitonic complexes for all investigated QDs remain in the range expected for the QDs without magnetic dopants \cite{nature2014, patton2003xx}. 

\begin{figure}[t]
\includegraphics[width=85mm]{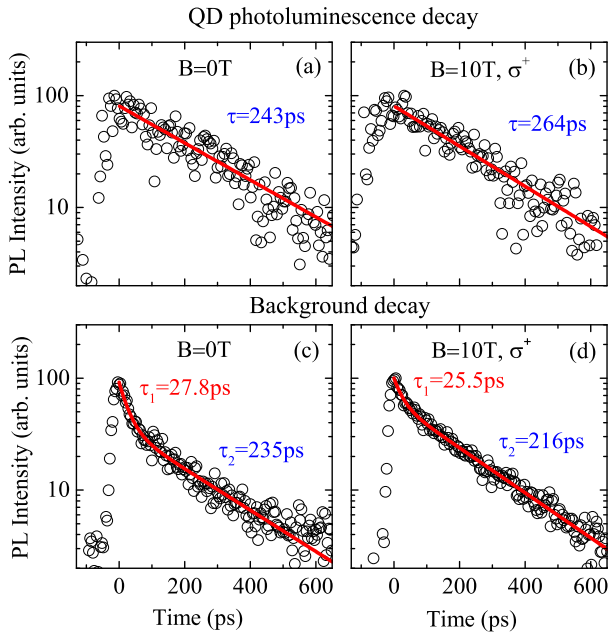}
\caption{(a), (b) Photoluminescence intensity decay of neutral exciton line related to an individual QD with a few Mn$^{2+}$ ions, after pulsed excitation in $B$ = 0 and $B$ = \SI{10}{T}. (c), (d) Analogous result obtained for the broad luminescence background present in the micro-PL measurements.}
\label{mikrotlo_dwuzanik}
\end{figure}

Similarly, in the case of the broad background no significant effect of the magnetic field on the PL decay was observed. The decay can be accurately described with two-exponential curve (see Fig. \ref{mikrotlo_dwuzanik}). The characteristic times determined from the fit of such curve are equal to \SI{230}{ps} $\pm$ \SI{30}{ps} for the longer component and about \SI{30}{ps} $\pm$ \SI{5}{ps} for the shorter component. Neither long nor short characteristic time reveals any significant dependence on both the magnetic field and circular polarization of detection.

All abovementioned results indicate that the PL quenching for excitons in QDs with a few Mn$^{2+}$ ions is not efficient. To detect the finest fingerprints of the quenching we need to be able to analyze the dynamics of individual spin states of the system. To distinguish emission lines related to those states we need to study a smaller system: a QD with an individual Mn$^{2+}$ ion. The neutral exciton luminescence spectrum of such a dot at zero magnetic field consists of 6 doubly degenerate emission lines\cite{cdtemn} corresponding to different projections of the  Mn$^{2+}$ ion spin onto to the quantization axis given by the exciton anisotropy axis (see Fig. \ref{symulacje0T}(a,f)).

\begin{figure}
\includegraphics[width=85mm]{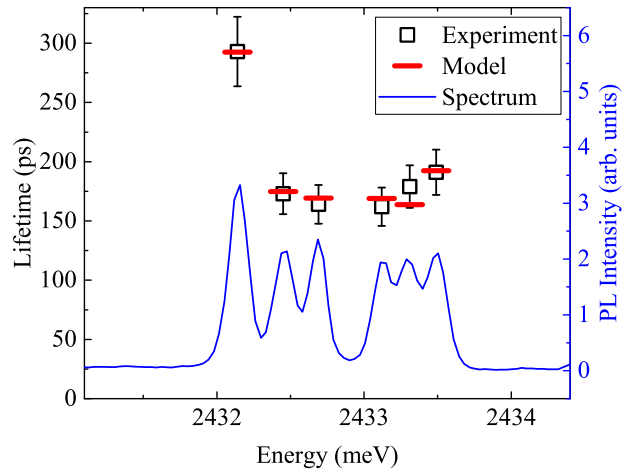}
\caption{PL spectrum of the neutral exciton from a QD with a single Mn$^{2+}$ ion. The points respresent the lifetimes determined for each out of six emission lines. The horizontal bars correspond to the predictions of the rate-equation model described in the text.}
\label{wynikimodel}
\end{figure}

We have performed time-resolved, high spectral resolution measurements of such a spectrum. Experimentally obtained lifetimes of the emission lines are presented in Fig. \ref{wynikimodel}. It is clearly visible that the lines characterized with the longest lifetime are those with extreme energies, i.e., related to the Mn$^{2+}$ states with extreme spin projections ($|S_{z}|=\frac{5}{2}$). Such a finding can be explained within a simple model, in which the luminescence decay of each of the exciton-ion states is modeled as a single-exponential decay with a rate corresponding to the sum of the rates corresponding to three mechanisms: radiative recombination channel and two non-radiative recombination channels, related either to the photoluminescence quenching or to the spin-flip relaxation of the exciton-ion complex occurring without the change of the ion spin. Since the rate of the radiative recombination should not depend on the spin projection of the Mn$^{2+}$ ion, the corresponding decay rate is assumed to be equal to $({\tau_\gamma}^{-1})$ for all six emission lines. As it was described in section III, the ion excitation is spin selective; only the middle four emission lines, corresponding to $|S_{z}|\le\frac{5}{2}$, should exhibit additional decay rate $({\tau_q}^{-1})$ due to this non-radiative recombination channel. Last factor, corresponding to the spin-flip of the exciton, will affect mostly the exciton interacting with the ion in the spin state $S_{z}=\pm\frac{5}{2}$, as the rate of this spin-flip significantly increases with the energy distance $\Delta E$ between the involved pair of exciton-ion states. In our model the spin-flip efficiency is assumed to be proportional to third power of energy difference (${\Delta E}^{3}$), in accordance with the theoretical predictions for a single-phonon-mediated spin relaxation of a bright exciton in a nonmagnetic QD~\cite{Tsitsishvili_PRB_2003}. Taking into account all aforementioned mechanisms, and assuming that the radiative recombination is much faster than both non-radiative processes, we obtain a simplified formula describing the decay rate of the exciton-ion complex in each out of six possible energy levels in singly-Mn-doped QD:

 \begin{gather}
 \label{decaymodel}
 \tau^{-1}= \tau_\gamma^{-1} + \tau _{a}^{-1} \left(\frac{S_z}{5/2}\right)^3 + 
\begin{cases}
  0 & S_z= \pm\frac{5}{2} \\    
  {\tau_q^{-1}} &S_z\neq\pm\frac{5}{2} .
\end{cases}
\end{gather}

By fitting our model to the experimental data we obtain a very good agreement between measured and calculated lifetimes of the individual emission lines (see Fig. \ref{wynikimodel}). This allows us to determine the values of the model parameters: $\tau_\gamma=\SI{232}{ps}$, $\tau_q=\SI{624}{ps}$ and $\tau_a=\SI{1124}{ps}$, which indicate that approximately one out of five excitons in a QD recombines non-radiatively exciting the Mn$^{2+}$ ion. On the other hand, it is about three out of four excitons that recombine non-radiatively in the previously discussed case of the wetting layer in the sample with QDs containing a few Mn$^{2+}$ ions (see Fig. \ref{mocbuly}). This brings us to the conclusion that the magnetic-ion-related non-radiative recombination channel of the excitons in the wetting layer is about 12 times more efficient than in the QD (when compared to the radiative recombination channel).

\begin{figure}
\includegraphics[width=85mm]{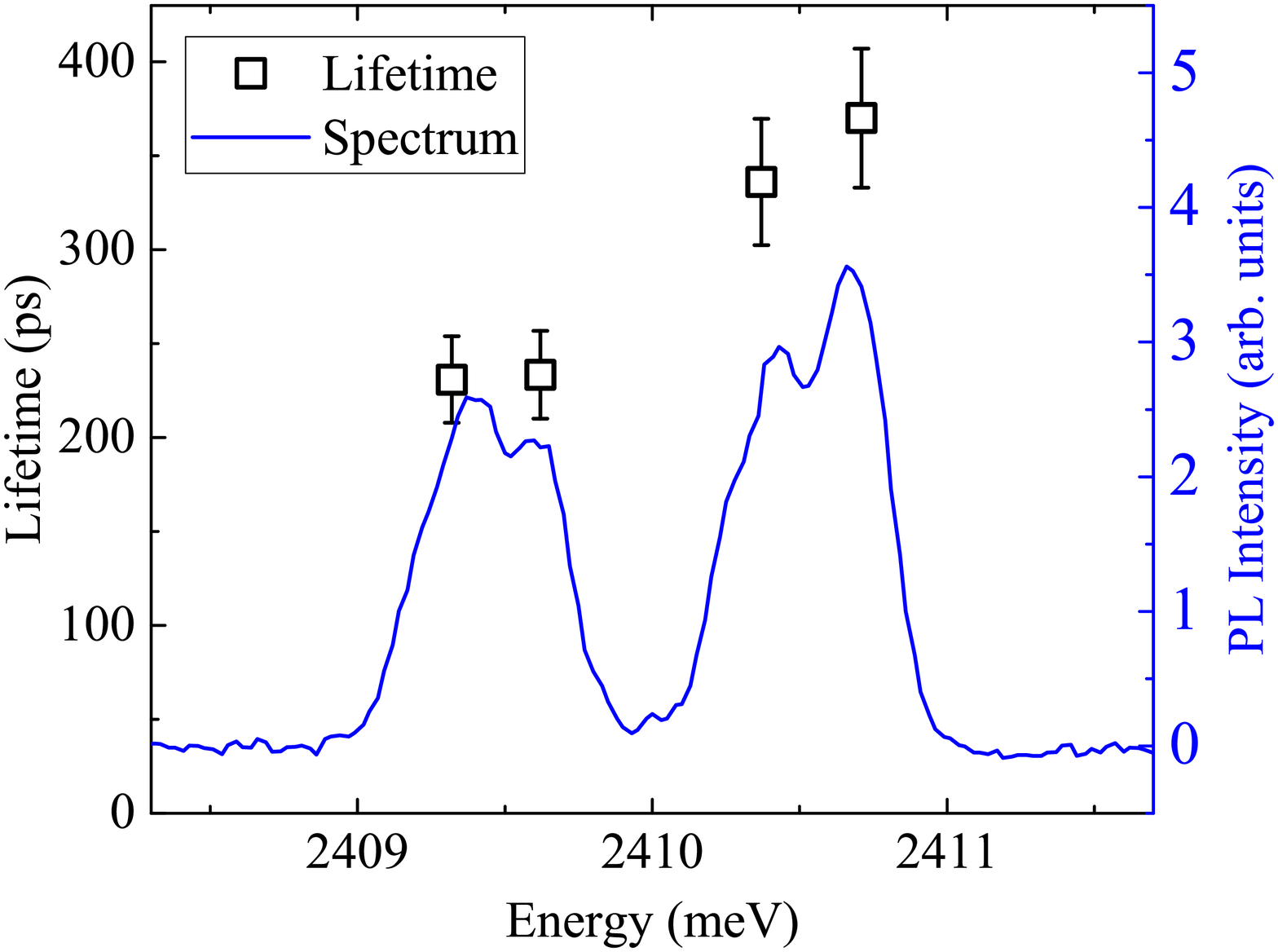}
\caption{PL spectrum of the negatively charged exciton from a QD with a single Mn$^{2+}$ ion with experimentally obtained luminescence lifetimes. The longest lifetime is observed for the highest energy line, for which non-radiative recombination involving the excitation of the Mn$^{2+}$ ion is spin-forbidden.}
\label{wynikimodelxp}
\end{figure}

One could argue that, within the experimental accuracy, the observed lifetimes of the neutral exciton emission lines (longest for the lowest-energy line) could be explained solely in terms of thermal relaxation of the exciton-ion system during the exciton lifetime. To make our explanation free of doubts, we have performed similar time-resolved analysis of the negatively charged exciton (X$^-$) in Mn-doped QD, the PL spectrum of which is shown in Fig. \ref{wynikimodelxp}. For X$^-$ the highest-energy line corresponds to the state of the X$^-$-Mn complex with the highest energy. At the same time, this line exhibits also the highest intensity, which cannot be explained by thermal relaxation. Instead, this finding can only be understood after taking into account non-radiative recombination channels involving the excitation of the Mn$^{2+}$ ion. In the case of the X$^-$ complex there are two such non-radiative channels. In order to analyze them, let us first focus on the case, in which the X$^-$ complex consists of the heavy hole with the total angular momentum projection on the growth axis $S^h_z = +3/2$. If the Mn$^{2+}$ ion becomes excited through non-radiative recombination of such a hole with an electron in $S^e_z = -1/2$ spin state, the selection rules are the same as for the neutral (bright) exciton, i.e., the spin of the Mn$^{2+}$ remains unchanged. As such, this process is possible only if the ion is in the $S_z = \pm 3/2$ or $S_z = \pm 1/2$ spin states before the excitation. The second possible scenario involves the excitation of the Mn$^{2+}$ ion by an electron-hole pair consisting of an electron with a spin projection ${J_e}_z = +1/2$. In such a case the selection rules are different, since the total angular momentum of both carriers is equal to $2$. Non-radiative recombination of such an electron-hole pair entails the change of the ion spin projection by 1 \cite{kossutdmsbook, chernenko2005, goryca2010PRB}. As such, this process is allowed only for the Mn$^{2+}$ ion occupying the spin states with $S_z=-5/2$, $S_z=-3/2$, or $S_z = \pm 1/2$ before the excitation. On this basis we deduce that both aforementioned non-radiative recombination channels are inactive only if the Mn$^{2+}$ ion is in the $S_z = +5/2$ spin state. Similar analysis of the case of the hole with an opposite angular momentum $J^h_z = -3/2$ leads to the conclusion that these channels remain inactive for the $S_z = -5/2$ spin state of the Mn$^{2+}$ ion. In both cases the only state of the X$^-$-Mn complex being not affected by the non-radiative recombination channel corresponds to the highest-energy line in the X$^-$ PL spectrum. As a consequence, this line is expected to exhibit the slowest decay dynamics. Our time-resolved spectroscopy confirms this prediction, as we observe significant prolongation of the lifetime as well as higher emission intensity for the higher-energy part of the X$^-$ PL spectrum, as seen in Fig. \ref{wynikimodelxp}.

Despite the above-described signatures of the excitonic PL quenching in a QD, the efficiency of this process is still very small as compared to the case of a bulk material doped with the Mn$^{2+}$ ions. A possible explanation of such a clear difference between these two cases might be related to fast saturation of non-radiative recombination channel, as the relaxation time of the Mn$^{2+}$ from its excited state can be very long, up to above ten microseconds\cite{beaulac2008mn}. In such a scenario, the Mn$^{2+}$ ion, once excited, persists in its excited state for a long time, and excitons subsequently injected to the QD cannot recombine non-radiatively. This mechanism, however, does not seem to apply in the physical situation which takes place in the studied samples. This is revealed, e.g., by the results of micro-PL measurements presented in Fig. \ref{mikrotlo}, which demonstrate the PL quenching to occur despite very high exciton injection rate, corresponding to about one exciton per Mn$^{2+}$ ion per few nanoseconds, that is a few orders of magnitude faster as compared to the Mn$^{2+}$ relaxation rate. The apparent lack of any saturation effects in our experiments can be explained by yet another mechanism, which was proposed recently in Ref. \onlinecite{bradshaw2012}. According to that theory, the Mn$^{2+}$ occupying its excited state can relax much faster transferring its energy to the exciton confined in a QD. Under an assumption of this process to be very efficient, it is conceivable to suppose that excited Mn$^{2+}$ ion may return to its ground state after injection of just one exciton to the dot. In such a case the non-radiative recombination channel of the exciton involving the Mn$^{2+}$ excitation cannot be saturated even for high excitation powers. This hypothesis is corroborated by the energy-structure of the X PL spectrum from a QD with a single Mn$^{2+}$ (see Fig.~\ref{wynikimodel}). As previously described, the invoked spectrum consists of six, approximately equidistant lines, which clearly indicates that each of these lines is due to the recombination of an exciton coupled to the Mn$^{2+}$ ion occupying the state with spin $S=5/2$. Such a condition is fulfilled only for the Mn$^{2+}$ ground state. This shows that the PL from the excitons interacting with the excited ion must be negligible, implying that the amount of time for which the Mn$^{2+}$ ion remains in such a state is very short, thus confirming the scenario of fast ion relaxation.

Altogether, the presented results show that although fingerprints of the spin-dependent PL quenching are visible in the QDs doped with single Mn$^{2+}$ ion, such quenching is not efficient for the excitons confined in QDs under investigation. However, it significantly affects the broad PL feature related to the wetting layer. The fact that the lifetime of this feature is not affected by the magnetic field suggests also that non-radiative recombination channels affect mostly higher (excited) excitonic states and not the relaxed states, from which the luminescence of the wetting layer come from.

\section{Conclusions}
We have performed a detailed magneto-optical study of the PL quenching in Mn-doped CdSe/ZnSe QDs. Our time-resolved and time-integrated experiments indicate that the magnetic-ions-related quenching does not occur effectively for the excitons confined in QDs, although the fingerprints of the expected spin-dependent quenching are visible. Nonetheless, all of these effects are still an order of magnitude weaker than the quenching effects observed in the PL experiments carried out on QD ensembles. Importantly, the quenching is found not to affect significantly the lifetime of the luminescence both for the QD and the wetting layer. As such, relatively efficient PL quenching observed in the latter case is thus most probably caused by non-radiative recombination channels of the excited photo-created carriers, before they reach the relaxed states.

\section*{Acknowledgments}
This work was partially supported by the Polish National Science Centre under decision DEC-2013/09/B/ST3/02603, DEC-2015/18/E/ST3/00559, DEC-2011/02/A/ST3/00131, DEC-2012/05/N/ST3/03209. T.S. was supported by the Polish National Science Centre through PhD scholarship Grant No. DEC-2016/20/T/ST3/00028. Project was carried out with the use of CePT, CeZaMat, and NLTK infrastructures financed by the European Union - the European Regional Development Fund within the Operational Programme ,,Innovative economy''.

\bibliographystyle{apsrev_my}

\end{document}